\documentclass[iop]{emulateapj}
\usepackage{natbib}
\usepackage{hyperref}
\hypersetup{
  colorlinks   = true,    
  urlcolor     = blue,    
  linkcolor    = blue,    
  citecolor    = blue      
}

\usepackage{graphicx}
\usepackage{float}

\slugcomment{Accepted for publication in AJ on 9/11/2018.}

\shorttitle{A deep search for planets around Vega}
\shortauthors{Meshkat et al.}

\begin{document}

\title{A deep search for planets in the inner 15 au around Vega}
\author{Tiffany Meshkat\altaffilmark{1,2}, Ricky Nilsson\altaffilmark{3,4}, 
Jonathan Aguilar\altaffilmark{5}, 
Gautam Vasisht\altaffilmark{2}, 
Rebecca Oppenheimer\altaffilmark{4}, 
Kate Y.L. Su\altaffilmark{6}, Eric Cady\altaffilmark{2}, Thomas Lockhart\altaffilmark{2}, Christopher Matthews\altaffilmark{2}, Richard Dekany\altaffilmark{3}, Jarron Leisenring\altaffilmark{6}, Marie Ygouf\altaffilmark{1}, Dimitri Mawet\altaffilmark{2,3}, Laurent Pueyo\altaffilmark{7}, Charles Beichman\altaffilmark{1,2}\\
}

\altaffiltext{1}{IPAC, California Institute of Technology, M/C 100-22, 1200 East California Boulevard, Pasadena, CA 91125, USA}
\altaffiltext{2}{Jet Propulsion Laboratory, California Institute of Technology, 4800 Oak Grove Drive, Pasadena, CA 91109, USA}
\altaffiltext{3}{Department of Astronomy, California Institute of Technology, 1200 East California Boulevard, MC 249-17, Pasadena, CA 91125, USA}
\altaffiltext{4}{Astrophysics Department, American Museum of Natural History, Central Park West at 79th Street, New York, NY 10024, USA}
\altaffiltext{5}{Department of Physics and Astronomy, The Johns Hopkins University, Baltimore, MD 21218, USA}
\altaffiltext{6}{Steward Observatory, University of Arizona, 933 N Cherry Avenue, Tucson, AZ 85721, USA}
\altaffiltext{7}{Space Telescope Science Institute, 3700 San Martin Drive, Baltimore, MD 21218, USA}

\begin{abstract}
We present the results of a deep high-contrast imaging search for planets around Vega. Vega is an ideal target for high-contrast imaging because it is bright, nearby, and young with a face-on two-belt debris disk which may be shaped by unseen planets. We obtained $J-$ and $H-$band data on Vega with the coronagraphic integral-field spectrograph Project\,1640 (P1640) at Palomar Observatory. Two nights of data were obtained in 2016, in poor seeing conditions, and two additional nights in more favorable conditions in 2017. In total, we obtained 5.5 hours of integration time on Vega in moderate to good seeing conditions ($<1\farcs5$). We did not detect any low mass companions in this system. Our data present the most sensitive contrast limits around Vega at very small separations (2--15\,au) thus far, allowing us to place new constraints on the companions which may be sculpting the Vega system. In addition to new constraints, as the deepest data obtained with P1640, these observations form the final legacy of the now decommissioned instrument.
\end{abstract}


\section{Introduction}
Nearly all directly imaged planets have been found around stars with bright circumstellar debris disks. These dusty disks contain grains down to a few microns in size, generated in collisional cascades of asteroids and comets \citep{Wyatt08}---bodies that are the remnants of planetesimals thought to be the building blocks of planet cores. This direct connection between debris disks and planets is seen in several of the currently known planetary systems (HR 8799; \citealt{Marois08,Marois10b}, $\beta$\,Pic; \citealt{Lagrange09,Lagrange10}, HD\,95086; \citealt{Rameau13b}, 51\,Eridani; \citealt{Macintosh15}), suggesting that debris disks may be signposts of exoplanetary systems \citep{Raymond11,Raymond12}. \citet{Meshkat17} perform a meta-analysis of new high-contrast imaging data supplemented by archival sensitivity limits to compare the occurrence rate of giant planets in dusty systems versus a well-defined control sample without dust belts under current detection limits. The occurrence of young giant planets around stars with debris disks is shown to be higher than those without debris disks at the 88\% confidence level, suggesting that these distributions are statistically distinct.
\smallskip

An additional hint for the presence of planets may be the signature of two temperatures in the debris disk's spectral energy distribution (SED): a warm inner belt and a cool outer belt. The dust-free gap between these belts may be caused by one or more planets accreting and rejecting the material as they form \citep{Chiang09,Quillen06}. Notably, the HR\,8799 planets, $\beta$\,Pic\,b, HD\,95086\,b, and 51\,Eridani\,b are in two-belt debris disk systems. 
\smallskip

Vega is one of the most well-studied stars in the Northern Hemisphere and an ideal target for high-contrast imaging searches. It is a relatively young \citep[$445\pm13$ Myr;][]{Yoon10} 2.5 $M_{\odot}$ A0V star, located $7.68\pm0.02$\,pc away \citep{vanLeeuwen07}. Vega is a bright (0th magnitude) star, making it favorable for good adaptive optics (AO) correction \citep{Metchev03} and thus deep contrast limits needed to detect the lowest mass planets. 

Vega has a vast, nearly face-on disk composed of small grains in the form of a disk halo first revealed by $Spitzer$ observations \citep{Su05}. JCMT 450 and 850\,$\mu$m images \citep{Holland17} reveal the smooth, axi-symmetric disk with a deconvolved fitted radius of the disk at 73 and 135\,au, respectively. 
These data also suggest that the center of the cold debris belt is offset from the star position by $2''$. This offset is smaller than previously detected in interferometric data at 1.3\,mm showing a peak offset of 8--14\arcsec\ \citep{Koerner01, Wilner02}. $Herschel$ data from 70 to 500\,$\mu$m are consistent with a smooth disk without the peak offset from the star \citep{Sibthorpe10}. $Herschel$ observations and re-analysis of the $Spitzer$ data reveal emission from an additional component of warm dust near the water ice line ($\sim$14\,au), spatially separated from the outer ($\sim$80\,au) cold belt \citep{Su13}. Based on the large gap between the inner warm and outer cold debris, and a companion mass limit from high-contrast imaging searches, \citet{Su13} suggest that the debris structures in Vega and its twin Fomalhaut system are signposts indicating multiple planets beyond the ice line. 

The face-on orientation of the Vega debris disk also makes it an optimal target for exoplanet imaging as a coplanar planet on a circular orbit will always be at the same angular separation from the star. In contrast, the face-on and fast rotating star makes radial velocity searches for planets extremely challenging. \citet{Janson15} combined deep $Spitzer$ observations with MMT observations \citep{Heinze08} to constrain planet masses in the Vega system, with an upper limit of $\sim$1--3\,$M_{\rm Jup}$ from 100 to 200 au from $Spitzer$ and $\sim$5--20\,$M_{\rm Jup}$ from 20 to 80\,au from MMT. 
The limits of $\sim$20\,$M_{\rm Jup}$ at 20\,au from \citet{Heinze08} represents the lowest-mass, innermost limits on Vega in the literature. 
\citet{Macintosh03} present wide-field ($>$50 au) Keck/NIRC2 $K$-band data with limits down to $\sim$5 Jupiter masses. Additionally, there is archival Keck/NIRC2 high-contrast imaging data on Vega using a coronagraph with a modest inner working angle of $1\arcsec$, thus not probing the very inner region of Vega, which remain largely unexplored. As part of the Lyot Project, \citet{Hinkley2007} present observations of the inner regions around Vega, on similar spatial scales to those reachable by P1640, achieving $H$-band contrasts with mass limits corresponding to 135 and 43$\,M_{\rm Jup}$ at $0\farcs5$ and $1\farcs0$, respectively.

In this work, we present the results of four nights of Project\,1640 (P1640) high-contrast imaging data in the $J+H$ bands. In Section \ref{sec:observations}, we discuss the observations and data reduction.
In Section \ref{sec:results}, we show the results of our data reductions, and also discuss the detection limits in the context of the Vega debris disk system and compare these limits with previous results.

\section{Observations and Data Reduction}
\label{sec:observations}

\begin{table}[h]
\scriptsize
\caption{Observing log for Vega Palomar/P1640 data.}
\centering
\begin{tabular}{c c c c}
\hline \hline
Observation dates & Number of & Integration time & Total integration \\
UT & cubes & per ramp (sec) & time (min) \\
\hline
2016 Aug 19 & 137 & 93.0 & 212.4  \\ 
2016 Aug 20 & 57 & 93.0 & 88.4  \\
2017 June 04 & 137 & 93.0 & 212.4 \\
2017 June 05 & 115 & 93.0 & 178.3 \\

\hline
\end{tabular}
\label{table:targets}
\end{table}

Vega observations were performed with the P1640 instrument at Palomar Observatory's 5.1-m Hale telescope over two nights in 2016 (Run 3201, PI: Meshkat) and two nights in 2017 (Run 3372, PI: Meshkat).
P1640 \citep{Soummer2009,Hinkley2009,Hinkley2011,Oppenheimer2012} is a coronagraphic integral-field spectrograph (IFS) with an internal wavefront sensing system \citep[CAL;][]{Zhai2012,Cady2013,Vasisht2014}, used in conjunction with the PALM-3000 \citep[P3k;][]{Dekany2013a} extreme AO system at the Hale telescope. The instrument covers a wavelength range of 969--1797\,nm, encompassing the near-IR $Y$, $J$, and $H$ bands in 32 spectral channels, at a spectral resolution of $\Delta\lambda = 26.7$\,nm, with a total field-of-view (FOV) of $\sim 3{\farcs}8 \times 3{\farcs}8$. 

As the Hale telescope sits on an equatorial mount, there is no field rotation that can be used for angular differential imaging \citep[ADI; see, e.g.,][]{Marois06}, as is common amongst other high-contrast imaging instruments. Instead, the IFS allows the wavelength dependent spatial distribution of speckle noise to be used for spectral differential imaging \citep[SDI;][]{Sparks02}, which can separate real astrophysical sources from the quasi-static speckles. Astrometric satellite spots, introduced by applying a fixed sinusoidal pattern on P3k's deformable mirror (DM), are used for determining the location of the star behind the occulting coronagraphic mask to sub-pixel precision, as well as for photometric calibration. Table \ref{table:targets} lists details about the observations for all four nights, including observation date, number of exposures, integration time per exposure, and total integration time.

\begin{figure}[t]
 \epsscale{1.0}
 \plotone{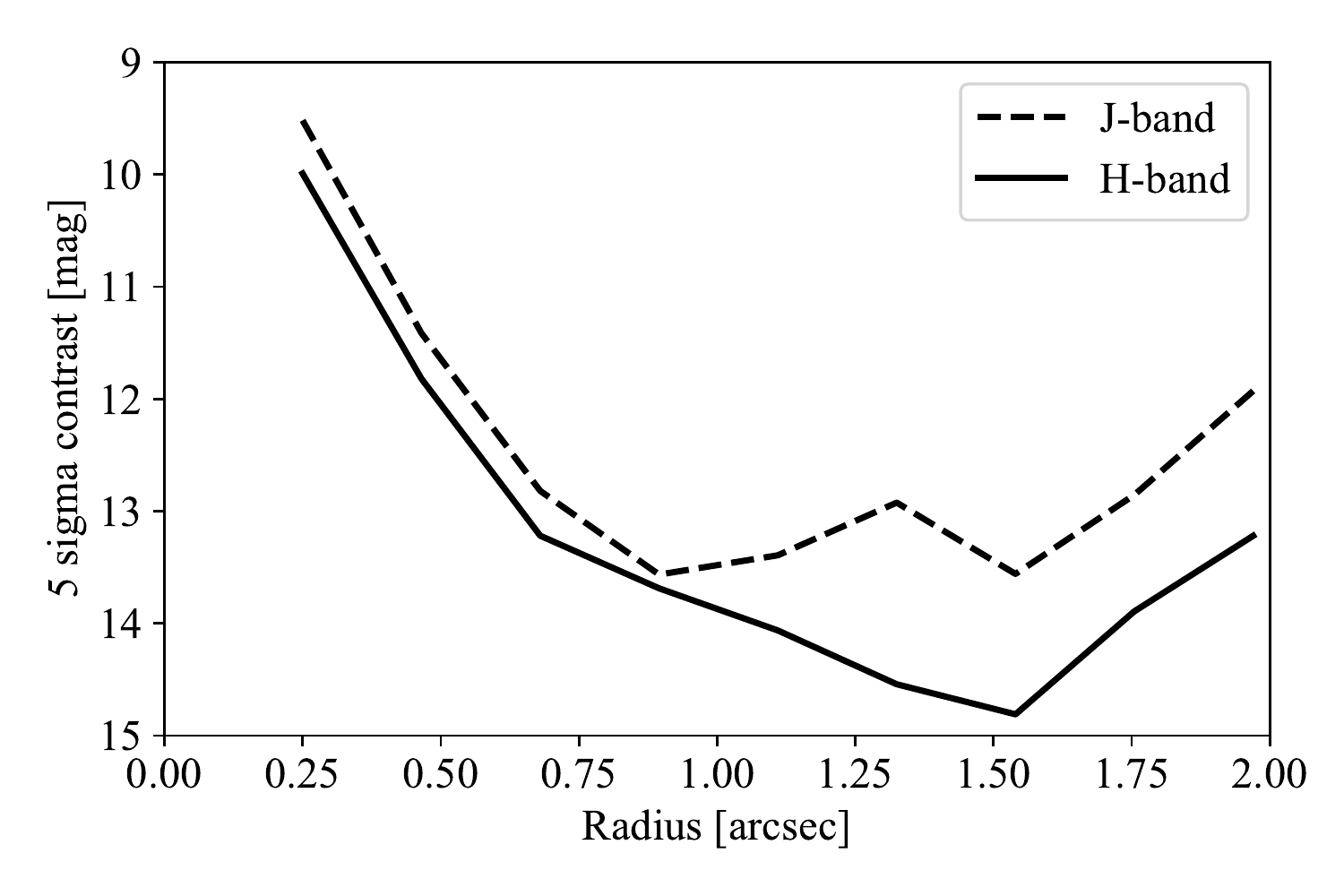}
 \caption{Contrast limits in magnitudes for $J$ (dashed line) and $H$ (solid line) band binned data. }
 \label{fig:contrast_curves}
 \end{figure}

\subsection{Data Reduction}

Raw IFS images were processed with the P1640 pre-processing pipeline PCXP \citep{Zimmerman11} to extract the 40,000 tightly packed spectra and produce data cubes ($x \times y \times \lambda$). Calibration laser exposures at 1310\,nm and 1550\,nm are combined with sky flats to create a focal plane solution that maps individual spectra to IFS lenslets and corresponding spaxel positions.

We generated post-processed data with two software packages: pyKLIP \citep{Wang18} and S4 \citep{Fergus14}. Both packages perform image registration using the astrometric satellite spots that track the location of the stellar PSF, which is centered behind the focal plane mask. pyKLIP was developed as an instrument-agnostic framework for processing ADI/SDI/ADI+SDI data with the KLIP PSF subtraction algorithm (\citealt{Soummer12, Amara12}), and has a P1640 instrument module for processing P1640 data. 
S4 was developed specifically for processing of P1640 data. The two packages are similar in that they use Principal Component Analysis (PCA) to capture the observed data variance and build a lower dimensional model of the data. However, while pyKLIP uses radially scaled images (where speckles remain stationary in radius-wavelength space) to model the spatial structure of the speckles, S4 models the joint spatial-spectral structure of the non-scaled data, conserving information about the morphology of the quasi-static speckle pattern in each channel. The derived model is subtracted from the data cubes, and the residual data can be inspected (by eye or with more sophisticated PSF matching techniques) to reveal stationary point sources. Each night was processed separately. No significant point sources were detected in the four separate nights of data.

We determined the detection limits achieved in these post-processed data using three different techniques for contrast estimation, in order to verify the derived accuracy of each method. The first is a detection limit pipeline developed for photometry known as optimized principal component analysis \citep[oPCA;][]{Meshkat14} which was adapted to IFS data using the pyKLIP post-processing package. The second is the pyKLIP detection limit pipeline, and the third is a full-field-of-view contrast estimator for S4 residual cubes. The first two pipelines utilize the satellite spots as photometric reference PSFs for fake companion injection. The average of the four satellite spots in each frame is used to create a fake companion which is injected in the frames before post-processing. The flux and position of the fake companion is scaled in order to determine the $5\sigma$ detection limit of the data in annuli at different angular separations. By injecting fake companions, we account for self-subtraction as a result of the post-processing PSF subtraction algorithms. The parameters used in the KLIP analysis were 5 modes, 5 annuli, and 3 subsections. In the following discussion we present the results from the oPCA pipeline, but we confirmed that these three techniques for determining contrast limits all yield consistent results.

Data for all four nights were processed separately, with a seeing limit cut-off of $1\farcs5$ that resulted in 37, 33, 122, and 21 cubes from each night respectively. The resulting signal-to-noise of a fake injected companion is measured in the average of the four post-processed cubes. In order to maximize the signal from a point source, we binned the IFS data into two ``bands'', roughly corresponding to $J-$ and $H-$ bands. 

\autoref{fig:contrast_curves} shows the contrast limits achieved on the average of the four nights using the oPCA pipeline for $J-$ and $H-$ bands, where contrast is the flux ratio between the star and a detectable point source.

\subsection{Deep observation sensitivity gain}

Our Vega observations were designed to investigate what imaging contrasts the instrument could deliver in a prolonged sequence ($\sim$10\,hr total exposure) on a bright star; note that the total exposure time over 4 nights was 11.5\,hr, out of which only 5.5\,hr was deemed usable given the selection criterion and the marginal seeing conditions. The P3k AO system delivers its best possible natural guide star performance on Vega, as the star allows both fast temporal sampling (1\,kHz) and high spatial sampling ($64 \times 64$ across the pupil) with negligible photon noise contribution to wavefront estimation \citep{Dekany2013a}. In median seeing conditions we expect a post correction wavefront of $\sim$130\,nm rms, corresponding to $J$- and $H$-band Strehl ratios of 0.65 and 0.78 respectively; the residual wavefront error is primarily dominated by the AO loop lag.

\begin{figure}
\epsscale{1.2}
\plotone{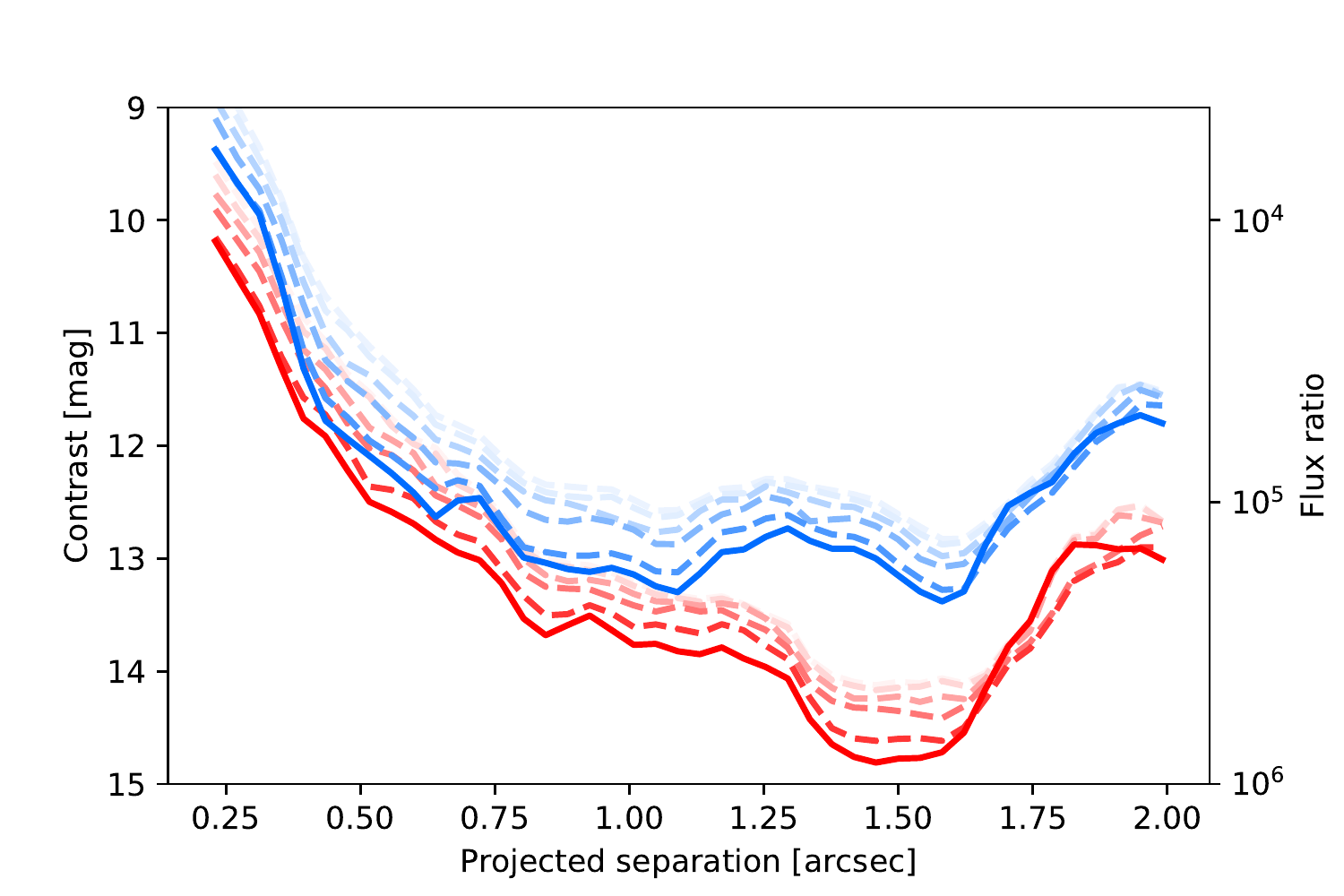}
\caption{Contrast curves for $J$- (blue) and $H$-band (red) for 100\% of the data (solid line) and 50\%, 25\%, 12.5\%, 6.3\%, 3.1\% in dashed faded lines.The 3.1\% dashed curve is the faintest curve. Contrast curves generated from subsets of data were computed with a moving average of data.}
\label{fig:contrast_subsets}
\end{figure}

\begin{figure}
\epsscale{1.2}
\plotone{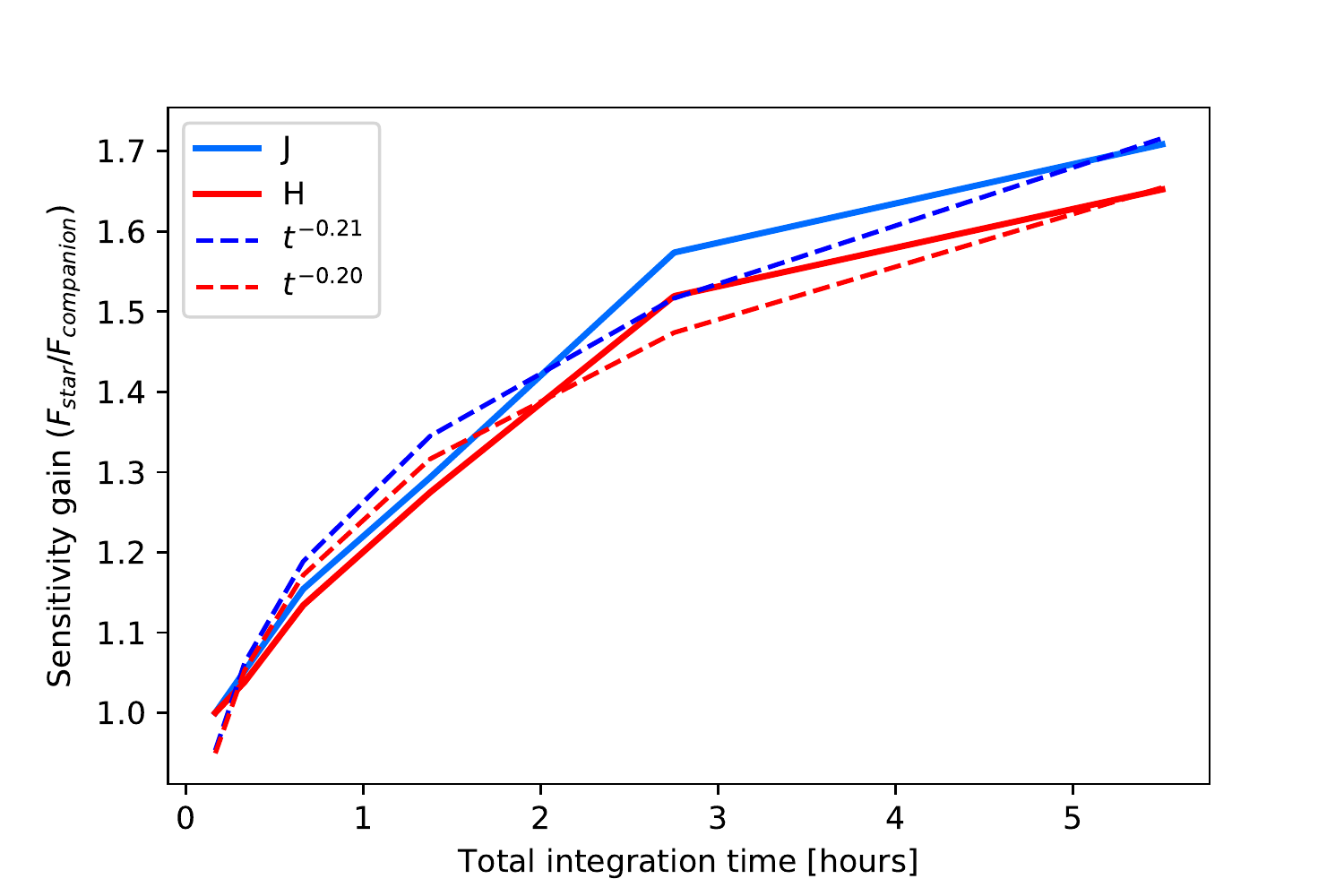}
\caption{Flux gain factor for $J$- (blue) and $H$-band (red) over the 5.5\,hr integration time. The flux gain is the average over all separations from \autoref{fig:contrast_subsets}. The dashed lines are best-fit power laws.}
\label{fig:gain}
\end{figure}

In the best case scenario, residual noise in post-processed images would be uncorrelated between images in an $n$-image sequence, except in the immediate vicinity of the star at $\leq 0\farcs5$ or within $\sim$3.8\,au, where SDI is less effective. In the $H$-band, where the system is optimized, co-adding $n$ images should then provide a near $n^{-1/2}$ improvement in detection gains; this translates to a factor of $\sim$14.6 or 2.9\,mag sensitivity over that in a single  image. In order to assess the actual gains achieved from co-adding, we create contrast curves for subsets of the data, from 3\% of the data to 100\% (only including data $< 1\farcs5$ threshold seeing). \autoref{fig:contrast_subsets} shows the contrasts achieved for 3.1\%, 6.3\%, 12.5\%, 25\%, 50\% and 100\% of the data in $J$ (blue) and $H$-band (red). The solid line indicates the curve with 100\% of the data included, and dashed lines are the subsets of data, with the faintest dashed line representing 3\% of the data. We measure the radial average of the gain in contrast with increasing subset sizes, shown in \autoref{fig:gain}. Our improvement from co-adding frames approximates a $n^{-1/5}$ power-law for both $J$- and $H$-band, shown as dashed lines.

Visual inspection of the post-processed images shows considerable residual structure surviving the standard filtering and PCA analysis. One clear source of noise is the low lying residual striping pattern due to the H2RG detector arising from temporal gain variations between readout channels. Another source is a band of diffracted light ascribed to a mosaic of under-responsive actuators on the high order deformable mirror (HODM). The changing influence function of these actuators, introduced by aging of the HODM, cause temporal and spatial effects on the contrast that are ill-understood. These structures clearly do not filter or average well, and degrade the azimuthal contrast and its averaging statistics.

A similar deviation from the naively expected $n^{-1/2}$ contrast improvement with exposure time in high-contrast AO coronagraphic imaging was demonstrated for the Lyot Project \citep{Hinkley2007}, albeit for smaller dynamic range. This is largely an effect of the Rician distribution of the speckle noise, which retains its non-Gaussian nature after imperfect speckle removal, but will get closer to $n^{-1/2}$ as processing techniques like those used by pyKLIP and S4 improve.

 \begin{figure}[t]
 \epsscale{1.2}
\plotone{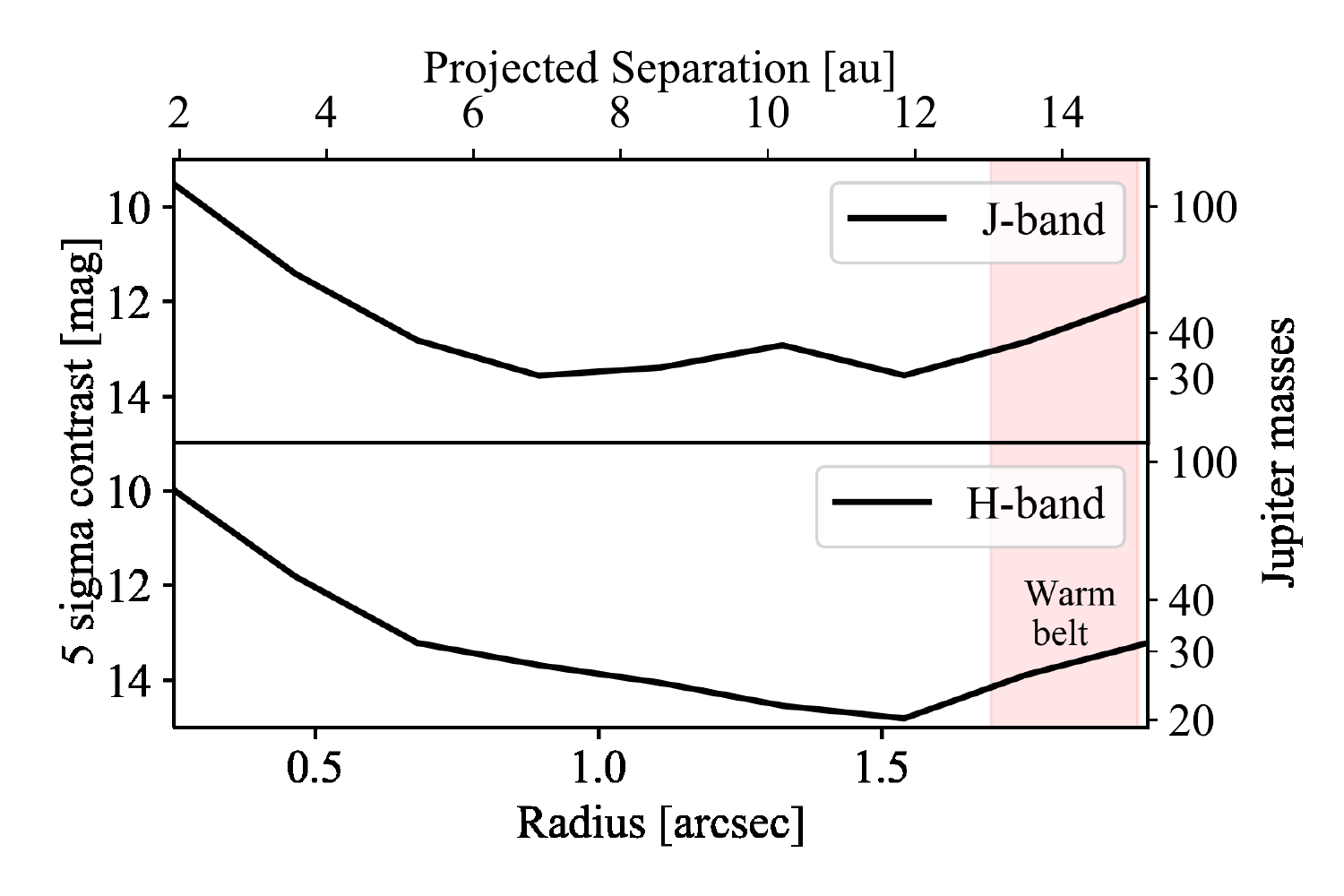}
 \caption{Mass limit plots for the binned $J$ (top) and $H$ (bottom) data. Masses are derived using the COND evolutionary model \citep{Chabrier00,Baraffe03}. The pink region from 13-15 au is the approximate location of the warm inner debris belt in the Vega system \citep{Su13}.}
 \label{fig:mass_limit}
 \end{figure}
 
\section{Results and Discussion}
\label{sec:results}

\subsection{Companion limits}
We convert our contrast limits to mass limits using the COND-AMES evolutionary model \citep{Chabrier00,Baraffe03}. The COND model is used for direct comparison with previous Vega analyses (i.e. \citealt{Janson15}). We note that the COND model is limited in that it presents an extreme where the dust opacity has been neglected, simulating a case where there is minimum dust content in the photosphere. At the magnitudes probed by our study, corresponding to effective temperatures of $\sim$2000-1000 K, this may not be completely appropriate, but we have adopted it as a conservative assumption that is consistent with previous studies. Using DUSTY models would change the minimum detectable mass to $\sim$10\,$M_{\rm Jup}$. \autoref{fig:mass_limit} shows our mass detection limits in $J-$ and $H-$bands, with the approximate location of the warm, inner debris belt marked as a pink region from approximately 13--15\,au, inferred from SED fitting \citep{Su13}. These limits allow us to rule out companions responsible for sculpting the inside of the warm debris belt at $\sim$12 au down to 20\,$M_{\rm Jup}$ for $H$-band and 30\,$M_{\rm Jup}$ for $J$-band.

\begin{figure}
\epsscale{1.2}
\plotone{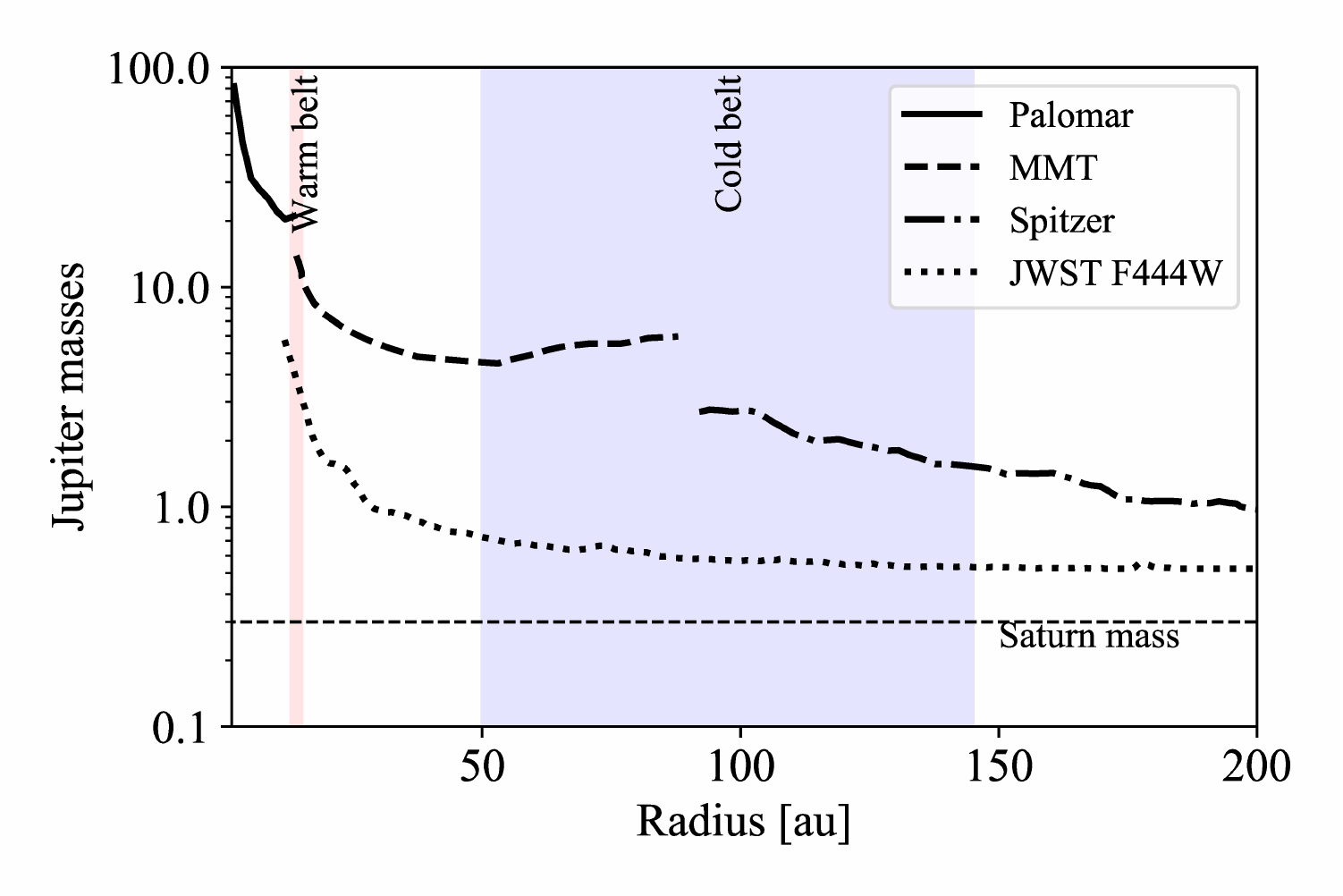}
\caption{Mass limits around Vega from 2 to 200 au, including the binned $H$-band Palomar/P1640 results presented in this work, as well as MMT \citep{Heinze08}, $Spitzer$ \citep{Janson15}, and predicted JWST data. All mass limit curves were estimated using the COND-AMES evolutionary model \citep{Chabrier00,Baraffe03}.} The debris disk inner and outer belt approximate locations are highlighted in the red and blue regions. The JWST curve reaches a lower mass floor at masses of 0.5\,$M_{\rm Jup}$. This mass limit is not physical but due to the mass limit of the COND-AMES model.
\label{fig:mass_combined}
\end{figure}

These data put limits on the low-mass stellar and brown dwarf companions which could be responsible for sculpting the Vega debris disks. We compare our results with previous contrast limits from \citet{Heinze08} with MMT and \citet{Janson15} with $Spitzer$ in \autoref{fig:mass_combined}. The combined results of these three datasets represents a complete limit of companions from 2 to 200\,au. The approximate locations of the warm inner and cold outer disk are labeled in red and blue, respectively.

\begin{figure}
\epsscale{1.0}
\plotone{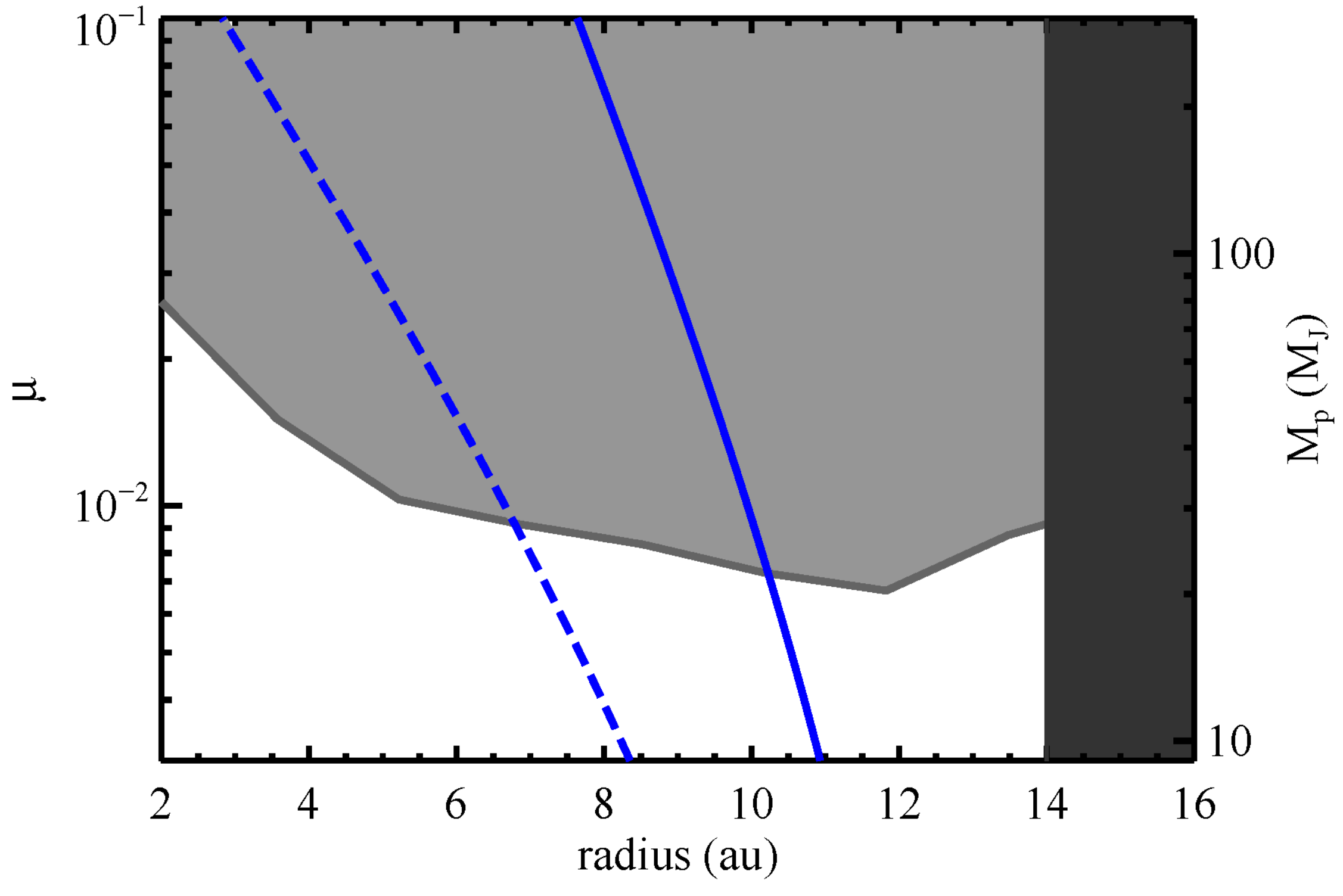}
\caption{The mass and location of possible circular, shepherding planets (shown as blue solid line) to maintain the inner edge (assumed at 14\,au) of the ice-line belt. The dark gray area marks the location of the inner belt, and the light gray area is the mass regime ruled out by our observation. The blue dashed line is the inner extension of the planet chaotic zone, detailed in Section \ref{sec:results}. }
\label{fig:inner_chaotic}
\end{figure}

\autoref{fig:inner_chaotic} shows the inner and outer chaotic zones interior to the warm inner belt, assuming that it has an inner boundary at 14\,au. We adopt the chaotic zone formula, numerically derived by \citet{Morrison15} designed for planets at circular orbits assuming high values of planet-to-star mass ratio ($\mu$). The solid blue line shows the semi-major axis with respect to planet mass for an outer chaotic zone that reaches the inner boundary of the inner belt at 14\,au, i.e., the likely location of a circular shepherding planet for a given mass to maintain the inner edge of the warm belt. 

\citet{Raymond14} perform dynamical simulations to constrain the masses of planets interior to the outer cold debris belt by analyzing planetesimal scattering. This analysis finds that low mass planets ($< 1\,M_{\rm Jup}$) orbiting at 5-10 au can be responsible for replenishing the hot exosodiacal dust of the inner belt, suggesting our data are not sensitive enough to place meaningful limits on these simulations.

\subsection{Debris disk limits}
We do not detect the Vega inner debris belt in our data, which is expected to be within our field of view based on SED modeling. Detection of the inner disk is particularly challenging because it is face-on. \citet{Absil06} detect a $1.29\pm0.19\%$ infrared excess relative to the Vega photosphere with FLUOR/CHARA in $K$-band. If the infrared excess is due to dust grains close to Vega ($< 10$\,au), the grains must be very small ($< 0.4\,\mu$m), have a fast blow-out time from radiation pressure, and thus requiring a very high replenishment rate. \citet{Defrere11} confirm this detection with IOTA/IONIC data in $H$-band, and suggest this may imply a late heavy bombardment-like event is occurring in the inner region around Vega. We convert our contrast limits to surface brightness limits, in order to determine if we can set limits on the amount of grains being blown out by radiation pressure. Adopting the total mass required for the $H$-band excess from \citet{Defrere11} and assuming 0.2\,$\mu$m silicate-like grains (albedo of 0.7 at $H$-band), the expected scattered light from these blow-out grains at $0\farcs5$ from the star is more than 2 orders of magnitude fainter than our contrast limit. Detecting the blowout small grains in scattered light will be extremely challenging.

\subsection{Future observations}
The planned JWST NIRCam GTO observations of Vega (PI Beichman, see \citealt{Beichman10}) will provide very deep coronagraphic data searching for companions beyond $1\farcs5$. \autoref{fig:mass_combined} demonstrates the deep sensitivity limits which are predicted to be achieved with the JWST NIRCam F444W filter. These data were generated using the python ETC and simulator for JWST NIRCam (pyNRC\footnote{https://pynrc.readthedocs.io}) which uses PSFs derived from WebbPSF \citep{Perrin14} paired with NIRCam's instrument performance to simulate a planned observation. The data presented here use the planned observation sequence for the GTO program with an exposure time of 1800 s, two roll positions, and assume a conservative wavefront error drift of 10\,nm between Vega and its linked reference target. Only simple reference subtraction was performed to produce the NIRCam contrast curves. More advanced post-processing techniques, such as PCA, should further improve detection limits interior to Vega's cold belt. The contrast curve is truncated shortward of $1\farcs5$ because Vega will saturate this inner region before the first read. We convert the contrast to masses using the AMES-COND model \citep{Chabrier00,Baraffe03}, for consistency with the other curves. The mass cut-off at 0.5\,$M_{\rm Jup}$ is not physical, but due to the mass lower limits in the AMES-COND model. Given the deep sensitivity limits, \citet{Beichman10} predict that JWST will likely place limits down to Saturn mass planets and contribute to our understanding of the gap between the inner and outer Vega debris belts.

\section{Conclusions}
\label{sec:conclusions}

We present the results of a deep search for companions around Vega with the Palomar P1640 high contrast imaging instrument. We combine data from several nights of P1640 integral-field spectrograph data spanning the $J + H$ band. We did not detect any point sources in our data. We present contrast curves and mass limits on Vega from our data (2--15\,au) and compare these with sensitivity limits with MMT, Spitzer, and predicted JWST limits. JWST data will provide a significant improvement over the previous data beyond 11\,au, in particular between the warm and cold debris belts. This is complemented by our P1640 sensitivity limits inside 10\,au and inside the warm dust belt.

\acknowledgments
We thank the anonymous referee for their helpful suggestions that improved this paper. We thank the Palomar mountain crew, especially Bruce Baker, Mike Doyle, Carolyn Heffner, John Henning, Greg van Idsinga, Steve Kunsman, Dan McKenna, Jean Mueller, Kajsa Peffer, Paul Nied, Joel Pearman, Kevin Rykoski, Carolyn Heffner, Jamey Eriksen, and Pam Thompson. We thank AAron Veicht for his contributions to the data acquisition during the observations. K.Y.L.S.  acknowledges  the  partial support  from  the  NASA  grant NNX15AI86G.

\bibliographystyle{apj}  

\end{document}